\documentclass[aps,prl,twocolumn,superscriptaddress,amsmath,amssymb]{revtex4-1}

\usepackage{graphicx}
\usepackage{subfigure}
\usepackage{multirow}
\usepackage{natbib}
\usepackage{array}
\usepackage{multirow,amssymb,amsbsy,amsmath,epstopdf}
\usepackage{color}
\usepackage{ulem}

\newcommand{\bra}[1]{\ensuremath{\left\langle #1\right|}}
\newcommand{\ket}[1]{\ensuremath{\left|#1\right\rangle}}
\usepackage[colorlinks,citecolor=blue]{hyperref}

\begin{document}
\title{Sharing Asymmetric Einstein-Podolsky-Rosen Steering with Projective Measurements}

\author{Yan-Xin Rong}
\affiliation{College of Physics and Optoelectronic Engineering, Ocean University of China, Qingdao 266100, People's Republic of China}

\author{Shuo Wang}
\affiliation{China Ship Research and Development Academy, Beijing, 100192 , People's Republic of China}

\author{Zhen-Fei Zhang}
\affiliation{College of Physics and Optoelectronic Engineering, Ocean University of China, Qingdao 266100, People's Republic of China}

\author{Yong-Jian Gu}
\affiliation{College of Physics and Optoelectronic Engineering, Ocean University of China, Qingdao 266100, People's Republic of China}

\author{Ya Xiao}\email{xiaoya@ouc.edu.cn}
\affiliation{College of Physics and Optoelectronic Engineering, Ocean University of China, Qingdao 266100, People's Republic of China}

\begin{abstract}
Recently, both global and local classical randomness-assisted projective measurement protocols have been employed to share Bell nonlocality of an entangled state among multiple sequential parties. Unlike Bell nonlocality, Einstein-Podolsky-Rosen (EPR) steering exhibits distinct asymmetric characteristics and serves as the necessary quantum resource for one-sided device-independent quantum information tasks. In this work, we propose a projective measurement protocol and investigate the shareability of EPR steering with steering radius criterion theoretically and experimentally. Our results reveal that arbitrarily many independent parties can share one-way steerability using projective measurements, even when no shared randomness is available. Furthermore, by leveraging only local randomness, asymmetric two-way steerability can also be shared. Our work not only deepens the understanding of the role of projective measurements in sharing quantum correlations but also opens up a new avenue for reutilizing asymmetric quantum correlations.
\end{abstract}

\maketitle

\section{Introduction}
Einstein-Podolsky-Rosen (EPR) steering describes the ability of one party, Alice, to nonlocally steer the state of the other party, Bob, even when Bob does not trust Alice. This property makes EPR steering a key resource for one-sided device-independent quantum key distribution~  \cite{Branciard2012,Opanchuk2014,Gehring2015,Walk2016}, randomness certification  \cite{Law2014,Passaro2015,Skrzypczyk2018}, quantum secret sharing \cite{Armstrong2015, Kogias2017, Wilkinson2023}  and quantum teleportation \cite{Reid2013_1, He2015}. EPR steering was first proposed by Schr\"odinger \cite{Schrödinger1935, Schrödinger1936} and then strictly redefined by Wiseman \textit{et al.} \cite{Wiseman2007}. In contrast to symmetric quantum entanglement and Bell nonlocality, EPR steering highlights an asymmetric property: the steerability from Alice to Bob may not be equal to  that of the opposite direction  and even allows for one-way steering, where Alice can steer Bob's state but not vice versa~\cite{Bowles2016}. This intrinsic asymmetry has potential applications  in asymmetric quantum network \cite{Cavalcanti2015, Wang2020}. Recently, asymmetric EPR steering has been experimentally demonstrated in both continuous and discrete variable systems \cite{Handchen2012,Sun2016, Wollmann2016, Xiao2017}. For a detailed introduction to EPR steering, please refer to the review article \cite{Uola2020}.

In the standard EPR steering tasks, each party performs basis projective measurements. Due to the monogamy constraint, no more than $N$ independent parties can share the steerability of an $N$-qubit state \cite{Reid2013_2}. In 2015, Silva \textit{et al.} demonstrated that this constraint can be removed if sequential unsharp (non-projective) measurements are adopted \cite{Silva2015}. They found that by properly modulating the measurement strength, the former party extracts enough nonlocality from the system while reserving enough for the next party, thereby enabling the sharing of Bell nonlocality of a two-qubit entangled state among multiple sequential parties \cite{Silva2015}. Since then, the sequential unsharp-measurement-based protocol has been widely employed to share EPR steering \cite{Sasmal2018, Shenoy2019, Choi2020, Han2021, Zhu2022, Han2022} and other types of quantum correlations, such as quantum entanglement \cite{Bera2018, Maity2020, Foletto2020_1, Srivastava2021, Pandit2022, Srivastava2022_1, Das2022, Srivastava2022_2, Hu2023, Li2023}, network nonlocality \cite{Hou2022, Halder2022, Mahato2022, Zhang2023, Mao2023} and quantum contextuality \cite{Kumari2019, Anwer2021, Chaturvedi2021, Kumari2023}. Additionally, these measurements have been demonstrated to yield outcomes that are unachievable in non-sequential scenarios, including unbounded randomness generation \cite{Curchod2017} and advantages in communication complexity \cite{Mohan2019, Foletto2020_2}. For more details, please refer to the review article on quantum correlation sharing \cite{Cai2024}.

However, implementing an unsharp measurement requires entangling the qubit with an auxiliary degree of freedom, which demands an increasingly complex experimental setup.  Recently, Steffinlongo \textit{et al}. proposed a protocol for sharing Bell nonlocality between multiple parties using only projective measurements \cite{Steffinlongo2022}. By allowing all involved parties to leverage global randomness and stochastically combine different projective measurement strategies, sequential violations of the Clauser-Horne-Shimony-Holt (CHSH) inequality can be witnessed. Later, Sasmal \textit{et al.} introduced a protocol for local randomness-assisted projective measurements that enables unbounded sharing of Bell nonlocality \cite{Sasmal2023}. These sharing protocols only require local projective measurements on qubits, which can be achieved using polarizers and quarter-wave plates in optical systems, making them experimentally easier \cite{Steffinlongo2022,Sasmal2023}.

On the experimental side, while numerous studies have demonstrated the feasibility of sharing Bell nonlocality through both sequential unsharp and projective measurement protocols \cite{Schiavon2017, Hu2018, Feng2020, Xiao2022}, investigations related to the sequential sharing of EPR steering  have been restricted to unsharp measurements \cite{Choi2020,Zhu2022}. Additionally, the number of parties that can share EPR steering is limited to four \cite{Choi2020}. Moreover, the authors employ a linear steering inequality to demonstrate the existence of steerability, but this criterion can not be used to verify whether the steerability is symmetric or asymmetric.

In this work, we extend Steffinlongo \textit{et al.}'s protocol and investigate the sharing of EPR steering using three different projective measurement strategies. The steerability is quantified by the two-measurement settings steering radius, which serves as a necessary and sufficient criterion. Our theoretical results show that one untrusted party can remotely steer all the states of an unbounded number of sequential parties simultaneously and independently. If each sequential party is allowed to leverage his private local randomness, the sharing of asymmetric two-way steering can also be observed. Experimentally, we take three parties as an example and show that it is possible to share both one-way steerability and asymmetric two-way steerability by employing only projective measurements.  Unlike Steffinlongo \textit{et al.}'s projective measurement protocol \cite{Steffinlongo2022}, our protocol requires only local randomness and negates the unitary operators of subsequent parties. Depart from Choi \textit{et al.}'s steering sharing work \cite{Choi2020}, we further demonstrate the feasibility of sharing asymmetric EPR-steering. Our work opens up new ways to reuse quantum correlations in asymmetric multi-party quantum information tasks.

\section{Theory and Scenarios}
\begin{figure}[!htp]
	\centering
	\includegraphics[width=1\linewidth]{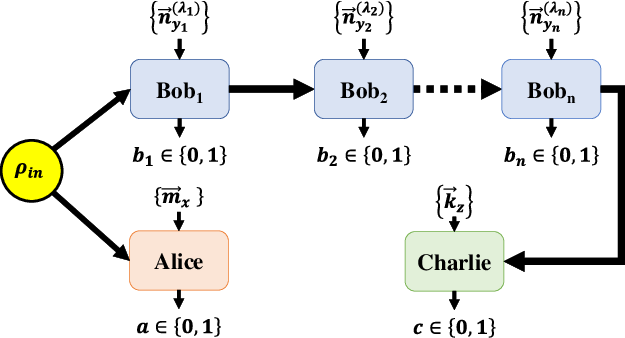}
	\caption{EPR steering sharing with local randomness-assisted projective  measurements. Initially, Alice and Bob$_1$ share a two-qubit entangled state $\rho_{in}$. Alice performs projective measurements on her qubit, while Bob$_1$ employs projective instruments on his qubit according to his local randomness ${\lambda_{1}}$, which follows the probability distribution $\{p(\lambda_1)\}$. Then Bob$_1$ relays the post-measurement state to Bob$_2$, who performs similar operations. This process continues until Charlie performs projective measurements on the received post-measurement state.}
	\label{model}		
\end{figure}

The EPR steering sharing scenario with projective measurements and local randomness is shown in FIG. \ref{model}. In this scenario, one particle of a two-qubit state $\rho_{in}$ is sent to Alice, while the other particle is subsequently distributed among multiple Bobs (Bob$_{i}$, $i\in \{1,2,...,n\}$) and Charlie, with Charlie being the last one. In this game, Alice's (each Bob's and Charlie's) task is to remotely steer the quantum state of each Bob's and Charlie's (Alice's) particle simultaneously and independently. Bob$_{i}$ or Charlie (Alice) will be convinced by Alice (Bob$_{i}$ or Charlie) if there is no local hidden state ensemble to construct his (her) conditional states. In the case of $N$-measurement settings, Alice (Charlie) performs the projective measurements of $N$ observables $A_{x}\equiv \{A_{a\vert x}=(I+(-1)^{a}\vec{m}_{x} \cdot \vec{\sigma})/2\}$  ($C_{z}\equiv \{C_{c\vert z}=(I+(-1)^{c}\vec{k}_{z} \cdot \vec{\sigma})/2\}$) according to the received measurement directions $\vec{m}_{x}\in \{\vec{m}_{1},\cdot\cdot\cdot, \vec{m}_{N}\}$ ( $\vec{k}_{z}\in \{\vec{k}_{1},\cdot\cdot\cdot, \vec{k}_{N}\}$), yielding binary output $a\in \{0, 1\}$ ($c\in \{0, 1\}$). $I$ is the identity matrix and $\vec{\sigma}=\{\sigma_{x}, \sigma_{y}, \sigma_{z}\}$ is the Pauli matrix. However, each Bob$_{i}$ employs a probabilistic projective instrument strategy by leveraging his local classical randomness $\lambda_i$, which is subjected to probability distributions $\{p(\lambda_i)\}$. For a certain $\lambda_i$, depending on the received measurement directions $\vec{n}^{(\lambda_i)}_{y_{i}}\in \{\vec{n}^{(\lambda_i)}_{1},\cdot\cdot\cdot, \vec{n}^{(\lambda_i)}_{N}\}$, Bob$_{i}$ applies a corresponding instrument characterized by Kraus operators $\{K^{(\lambda_{i})}_{b_{i}\vert y_{i}}\}$ to his particle. This produces a classical binary outcome $b_{i}\in \{0, 1\}$ recorded by Bob$_{i}$, and a qubit post-measurement state is relayed to the nearest Bob$_{i+1}$. This process continues until the post-measurement state is measured by the final Charlie. Notably, since the instrument of Bob${_i}$ realizes a projective measurement, the corresponding Kraus operators must satisfy the completeness relation $\forall y_{i}$, $B^{(\lambda_{i})}_{0\vert y_{i}}+B^{(\lambda_{i})}_{1\vert y_{i}}=I$, where $ B^{(\lambda_{i})}_{b_{i}\vert y_{i}}= K^{\dagger  (\lambda_{i})}_{b_{i}\vert y_{i}}K^{(\lambda_{i})}_{b_{i}\vert y_{i}}=[I+(-1)^{b_i}\vec{n}^{(\lambda_i)}_{y_{i}}\cdot\vec{\sigma}]/2$  are the corresponding elements of the projective measurement observable $B^{(\lambda_{i})}_{ y_{i}}=B^{(\lambda_{i})}_{0\vert y_{i}}-B^{(\lambda_{i})}_{1\vert y_{i}}$.  As sequential Bobs and Charlie are assumed to act independently, the state $ \rho_{AB_{i+1}}(\rho_{AC}) $ shared between Alice and Bob$_{i+1}$ (Charlie) can be described by the L\"uders rule \cite{Werner1989}:

\begin{equation}
	\begin{split}
		&\rho_{AB_{i+1}}= \frac{1}{2}\sum_{\lambda_i, y_i, 	b_i}p(\lambda_i) (I \otimes  K_{b_i\vert y_i}^{(\lambda_i)})\rho_{AB_{i}} (I \otimes K^{\dagger (\lambda_i)}_{b_i\vert y_i}),\\
		&\rho_{AC} = \frac{1}{2}\sum_{\lambda_n, y_n, 	b_n}p(\lambda_n) (I\otimes  K_{b_n\vert y_n}^{(\lambda_n)})\rho_{AB_{n}} (I \otimes K^{\dagger (\lambda_n)}_{b_n\vert y_n}),
	\end{split}
	\label{luders}
\end{equation}
where $ \rho_{AB_{1}}=\rho_{in} $. If Bob$_{i}$ (Charlie) performs measurement $B^{(\lambda_{i})}_{b_{i}\vert y_{i}}$ ($C_{c\vert z}$), the conditional state of Alice can be expressed as $\rho^{A}_{b_{i}\vert \vec{n}_{y_i}^{\lambda_{i}}}= {\rm Tr}_{B_{i}}\left[B^{(\lambda_{i})}_{b_{i}\vert y_{i}}\rho_{AB_{i}}\right]$ ($ \rho^{A}_{c\vert \vec{k}_{z}}= {\rm Tr}_{C}\left[C_{c\vert z}\rho_{AC}\right]$). Similarly, if Alice performs measurement $A_{a\vert x}$, the conditional state of Bob$_{i}$ (Charlie) can be expressed as $\rho^{B_{i}}_{a\vert \vec{m}_{x}}= {\rm Tr}_{A}\left[A_{a\vert x}\rho_{AB_{i}}\right]$ ($\rho^{C}_{a\vert \vec{m}_{x}}= {\rm Tr}_{A}\left[A_{a\vert x}\rho_{AC}\right]$).

Here, we adopt steering radius $R_{AB_{i}}$ to quantify the steerability from Alice to Bob$_{i}$, which serves as a necessary and sufficient criterion. It can be defined as \cite{Bowles2016,Xiao2017}
\begin{equation}\label{sd}
	R_{AB_{i}}=\max\limits_{\{\vec{m}_1,\cdot\cdot\cdot, \vec{m}_{N}\}} \lbrace \min \limits_{\lbrace   p^B_{i}\rho^B_{i}\rbrace} \lbrace {\rm max} \lbrace \vert  \vec{v}^B_{i} \vert \rbrace \rbrace \rbrace. 
\end{equation}
$\lbrace p^B_{i}\rho^B_{i}\rbrace$ is the local hidden state ensemble of Bob$ _{i} $, $\lvert \vec{v}^B_{i} \rvert$ is the length of the Bloch vector of the corresponding local hidden state $\rho^B_{i}$. If $ R_{AB_{i}}>1 $,  there is no local hidden state ensemble to construct the conditional states of Bob$_i$, which indicates that the steering task from Alice to Bob$_{i}$ is successful. Otherwise, the steering task fails if $R_{AB_{i}}\leq 1 $.  The steering radius for the case in which Alice steers Charlie's state ($R_{AC}$), Bob$_{i}$  steers Alice's state ($R_{B_{i}A}$), Charlie steers Alice's state ($R_{CA}$) can be defined in a similar way.

Consider a family of two-qubit states \cite{Bowles2016} 
\begin{equation}
	\rho(W,\theta) = W \ket{\psi(\theta)} \bra{\psi(\theta)} + \frac{(1-W)}{2} I \otimes \rho_{B}(\theta),
	\label{state}
\end{equation}
where $W\in[0, 1]$ and $\theta \in [0,\pi/2]$. $\ket{\psi(\theta)}\! = \! \cos{\theta} \! \ket{HH} \! + \! \sin{\theta} \! \ket{VV}$,  $\ket{H}$ and $\ket{V}$ represent the horizontal and vertical polarization, respectively. $\rho_{B}(\theta) = \mathrm{Tr}_A \left[ {\ket{\psi(\theta)} \bra{\psi(\theta)}} \right]$. Without loss of generality, we mainly concentrate on the simplest scenario involving three parties, namely Alice, Bob, and Charlie. In this scenario, only Bob employs local randomness-assisted projective instruments. For two-measurement settings, based on the symmetrical property of the steering ellipsoid of state $\rho(W,\theta) $ \cite{Jevtic2014,Jevtic2015}, the optimal measurement directions are $\{\vec{x}, \vec{z} \}$ for both steering directions. Therefore, with Alice's and Charlie's measurement observables respectively selected as $\{A_{0}=\sigma_{x}$, $A_{1}=\sigma_{z}\}$ and $\{C_{0}=\sigma_{x}$, $C_{1}=\sigma_{z}\}$, the measurement strategies of Bob are as follows: 

\textbf{Case 1 ($\lambda=1$): Both measurements are basis projections.} Bob measures $B^{(1)}_0=\sigma_x$ and $B^{(1)}_1=\sigma_z$. In this case, $R^{(1)}_{AB} \in [ 0, \sqrt{2} ]$, $R^{(1)}_{BA} \in [0, \sqrt{2}]$, $R^{(1)}_{AC} \in [0, 1/\sqrt{2}]$ and $ R^{(1)}_{CA} \in [0, 1] $. It is clear that the steering between Alice and Bob is permissible, however, the steering between Alice and Charlie is forbidden.

\textbf{Case 2 ($\lambda=2$): Both measurements are identity projections.} Bob measures $B^{(2)}_0 \!=\! B^{(2)}_1 \!=\! I$.  In this case, $R^{(2)}_{AB} \!=\! R^{(2)}_{AC} \! \in \! [0, \sqrt{2}]$, $R^{(2)}_{BA} \! \in \! [0,\sqrt{2}]$ and $R^{(2)}_{BA} \! \in \! [0,1]$. Thus, only the steering from Bob to Alice is forbidden. 

\textbf{Case 3 ($\lambda=3$): One measurement is an identity projection and the other is a basis projection.} Bob measures $B^{(3)}_0=I$ and $B^{(3)}_1=\sigma_z$. In this case, $R^{(3)}_{AB} \in [0, \sqrt{2}]$, $R^{(3)}_{BA} \in [0,1]$, $R^{(3)}_{AC} \in [0, \sqrt{5}/2]$ and $R^{(3)}_{CA} \in [0,\sqrt{5}/2]$. Similar to case 2, only the steering from Bob to Alice  is forbidden.

Obviously, the one-way steering can be shared with the measurement strategy shown in case 2 or case 3. But the sharing of two-way steering is impossible when only employing a deterministic measurement strategy.  If the stochastic combination of the above three deterministic measurement strategies is permitted, the sharing of two-way steering can be further achieved. This analysis is detailed in the Appendix I.

\section{Experimental setup and results}

\begin{figure}[htbp]
	\centering
	\includegraphics[width=\linewidth]{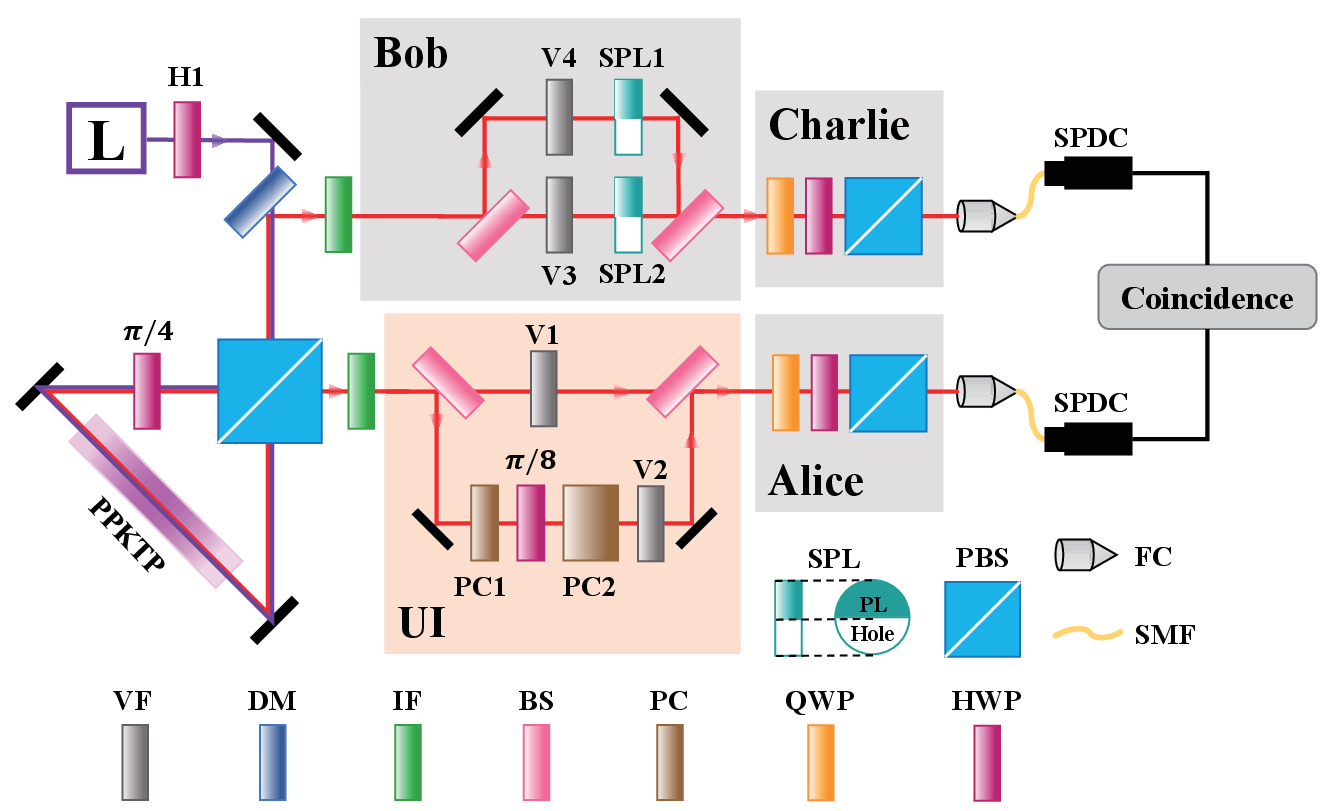}
	\caption{Experimental setup. A pair of polarization-entangled photons in the state $\ket{\psi(\theta)}$ is generated by pumping a type-II cut PPKTP crystal located in a Sagnac interferometer. After being filtered by interference filters (IFs), one of the down-conversion photons is directed to an unbalanced interferometer (UI) to further prepare the state described in Eq. (\ref{state}). Then, it is sent to Alice. Meanwhile, the other photon is sequentially sent to Bob and Charlie. A half-wave plate (H1) and two variable filters (V1 and V2) are respectively used to adjust the state parameters $\theta$ and $W$. Bob's projective  measurements can be simulated by two semi-circular polarizers (SPL1 and SPL2). For some appropriate axes, SPL1 and SPL2 can act as basis projections of $\sigma_{x}$, $\sigma_{z}$ and the identity projection $I$. Two variable filters (V3 and V4) are used to adjust the probability distribution ${p(\lambda)}$. Alice and Charlie employ quarter-wave plates (QWPs), half-wave plates (HWPs), and polarization beam splitters (PBSs) to perform the associated projective measurements.}   
	\label{setup}
	\vspace{-1em}
\end{figure}

FIG. \ref{setup} shows the experimental setup. A 10 mm long periodically poled potassium titanyl phosphate (PPKTP) crystal, located within a  Sagnac interferometer, is pumped both clockwise and counterclockwise by a 405 nm continuous diode laser to generate a two-qubit polarization-entangled photon state $\ket{\psi(\theta)}$. A half-wave plate (H1) is used to adjust the state parameter $\theta $. The pump beam is filtered by two interference filters (IFs). One of the down-conversion photons is directed to an unbalanced interferometer (UI), where a beam splitter (BS) separates it into two paths. In the upper path, the state remains unchanged, but in the lower path, two birefringent crystals (PC1, PC2) with sufficiently long lengths entirely destroy the coherence between $\ket{H}$ and $\ket{V}$.  Combining these two paths enables the preparation of arbitrary two-qubit states in the form described by Eq. (\ref{state}). The parameter $W$ can be adjusted by rotating two variable filters (V1 and V2). After passing through the UI, the photon is received by Alice, who uses a half-wave plate (HWP), a quarter-wave plate (QWP), and a polarization beam splitter (PBS) to implement her desired projective measurements.

The other down-conversion photon is first sent to Bob. We consider a scenario where Bob can randomly combine no more than two different projective measurement strategies. This can be realized by the other UI. A semi-circular polarizer (SPL) in each path allows Bob to implement the measurement strategies shown in cases 1 to 3. Specifically, as the photon passes through the hole part,  SPL  behaves as the identity projection $I$. As the photon traverses the polarizer section,  SPL  behaves as $\sigma_{x}$ basis measurements when its axis is set at $-\pi/4$ and $\pi/4$, and as $\sigma_{z}$ basis measurements when its axis is set at $0$ and $\pi/2$. The probability distribution ${p(\lambda)}$ can be adjusted using variable filters (V3 and V4). After combining these two paths, the photon is sequentially sent to Charlie, who performs the desired projective measurements using the same setup as Alice. By evaluating the respective steering radius, we can check whether the state between Alice and Bob, as well as between Alice and Charlie, is two-way steerable, one-way steerable, or unsteerable.

\begin{figure}[!ht]
	\centering
	\includegraphics[width=\linewidth]{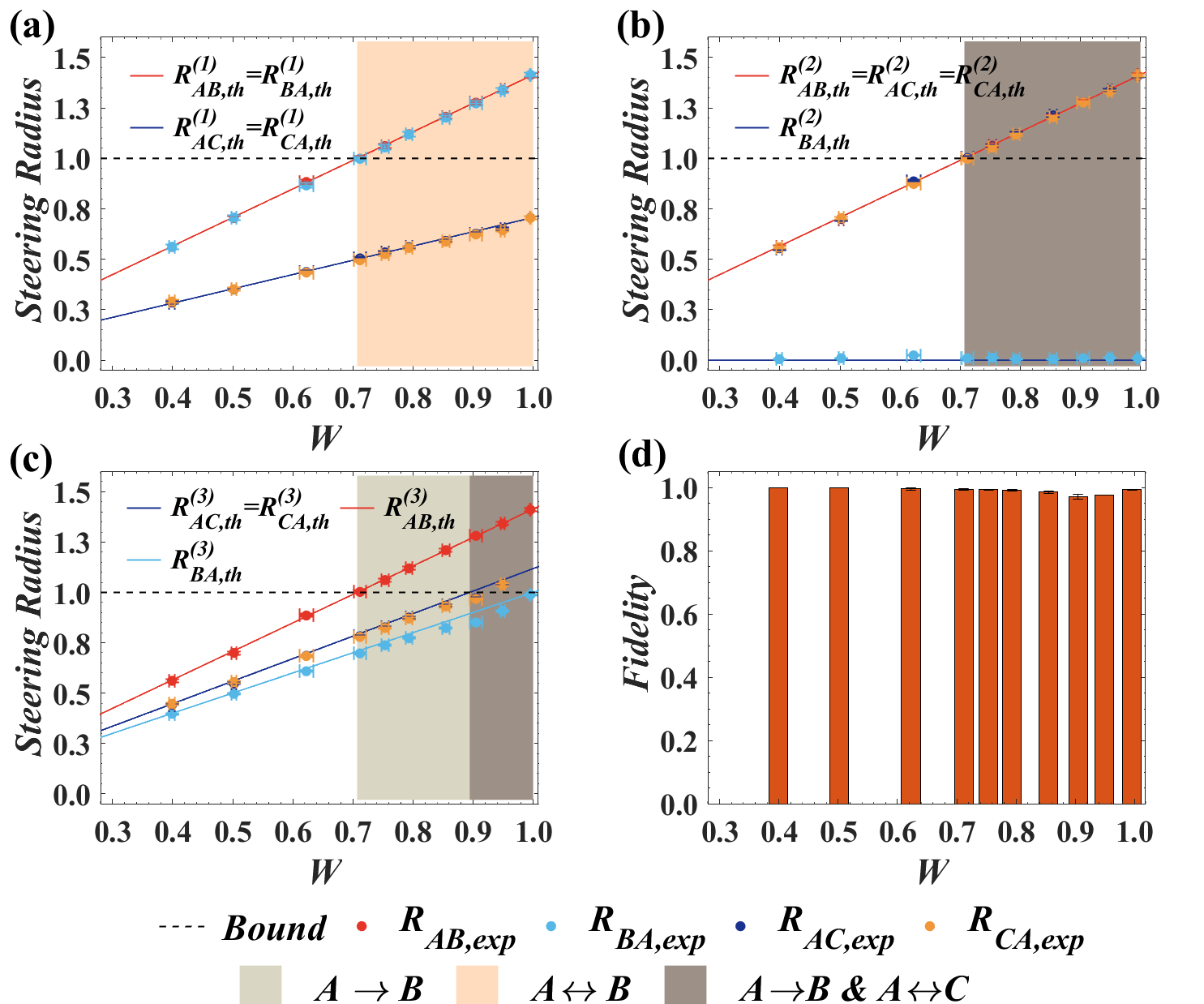}
	\caption{The steerability of a family of symmetric states with $\theta=\pi/4$.  Steering radii for the measurement strategies shown in (a). case 1, (b). case 2 and (c). case 3. Theoretical and experimental results are depicted by solid lines and circles, respectively. The dashed black lines denote boundaries between steerable and unsteerable states. The vertical colored regions represent different direction-related steering classes: one-way steerable from Alice to Bob ($A \to B$), two-way steerable between Alice and Bob ($A \leftrightarrow B$), one-way steerable from Alice to Bob and two-way steerable between Alice and Charlie ($A \rightarrow B \& A \leftrightarrow C$). (d). The fidelity of the experimental states shown in (a-c).  Error bars are estimated by the Poissonian statistics of two-photon coincidences.}
	\label{result1}
\end{figure}

We first investigate the sharing of EPR steering with a family of symmetric states (Werner states) by setting $\theta \!=\! \pi/4$. The experimental results for ten states with $W =\{0.3990 \!\pm\! 0.0050, 0.5020\!\pm\!0.0039, 0.6225\!\pm\!0.0113, 0.7118\!\pm\!0.0107, 0.7534\!\pm\!0.0063, 0.7932\!\pm\!0.0053, 0.8547\!\pm\!0.0058, 0.9048\!\pm\!0.0108, 0.9493\!\pm\!0.0037, 0.9954\!\pm\!0.0007 \}$ under three deterministic projective strategies are depicted in FIG. \ref{result1}. As shown in FIG. \ref{result1} (a-c), the steering radii exhibit linear behavior with respect to  $W$. It is evident that in case 1, the steering radius from Alice to Bob (Charlie) is equal to that from Bob (Charlie) to Alice. Two-way steering between Alice and Bob appears when $W \!\in\! (1/\sqrt{2},1]$. However, the sequential sharing of EPR steering is absent. In case 2, one-way steering from Alice to Bob and two-way steering between Alice and Charlie are allowed in the same range $W \!\in\! (1/\sqrt{2},1]$. Clearly, Alice can simultaneously steer the states of Bob and Charlie. By employing the measurement strategy shown in case 3, the range in which Alice can simultaneously steer Bob and Charlie decreases to $W \!\in\! (2/\sqrt{5},1]$. Unlike the previous two cases, an asymmetric steering range ($W \!\in\! (1/\sqrt{2},2/\sqrt{5})$) emerges, within which only Alice can steer Bob's state. The corresponding state fidelities are further shown in  FIG. \ref{result1} (d), whose average value is about $0.9895 \!\pm\! 0.0091$. 

\begin{figure}[!ht]
	\centering
	\includegraphics[width=\linewidth]{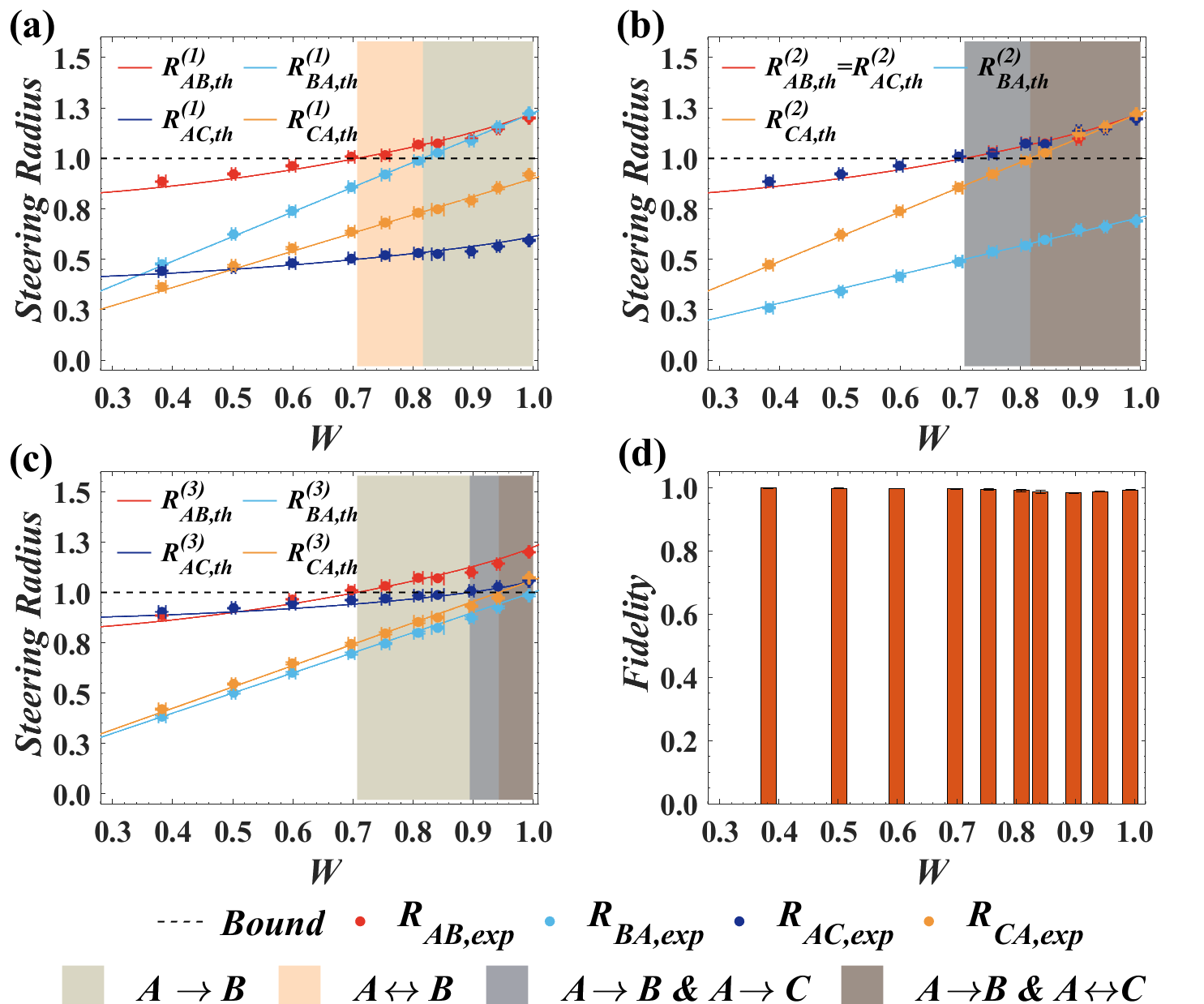}
	\caption{The steerability of a family of asymmetric states with $\theta=\pi/8$.  Steering radii for the measurement strategies shown in (a). case 1, (b). case 2 and (c). case 3. Theoretical and experimental results are depicted by solid lines and circles, respectively. The dashed black lines denote boundaries between steerable and unsteerable states. The vertical colored regions represent different direction-related steering classes: one-way steerable from Alice to Bob ($A \to B$), two-way steerable between Alice and Bob ($A \leftrightarrow B$), one-way steerable from Alice to both Bob and Charlie ($A \to B \& A \to C$), one-way steerable from Alice to Bob and two-way steerable between Alice and Charlie ($A \to B \& A \leftrightarrow C$). (d). The fidelity of the experimental states is shown in (a-c).  Error bars are estimated by the Poissonian statistics of two-photon coincidences.}
	\label{result2}
\end{figure}

The experimental results corresponding to those above three deterministic projective strategies with a family of asymmetric states ($\theta\!=\!\pi/8$) are shown in FIG. \ref{result2}. Compared with the previous scenario of symmetric states, each deterministic projective strategy reveals more direction-related steering classes. Specifically, one-way steering from Alice to Bob is further allowed when $W \! \in \! (1/\sqrt{2},\sqrt{2/3})$ in case 1. Additionally, in case 2, one-way steering from Alice to both Bob and Charlie appears within the same range $(W \! \in \! 1/\sqrt{2},\sqrt{2/3})$. In case 3, similar changes are also observed: Alice can simultaneously steer the states of Bob and Charlie, but not vice versa, when $W\in(2/\sqrt{5},\sqrt{2/3})$. Obviously, the one-way steering can be shared among Alice, Bob, and Charlie with the measurement strategies shown in case 2 and case 3. The corresponding state fidelities are further shown in FIG. \ref{result2} (d), whose average value is about $0.9918 \pm 0.0050$.

\begin{figure}[ht]
	\centering
	\includegraphics[width=\linewidth]{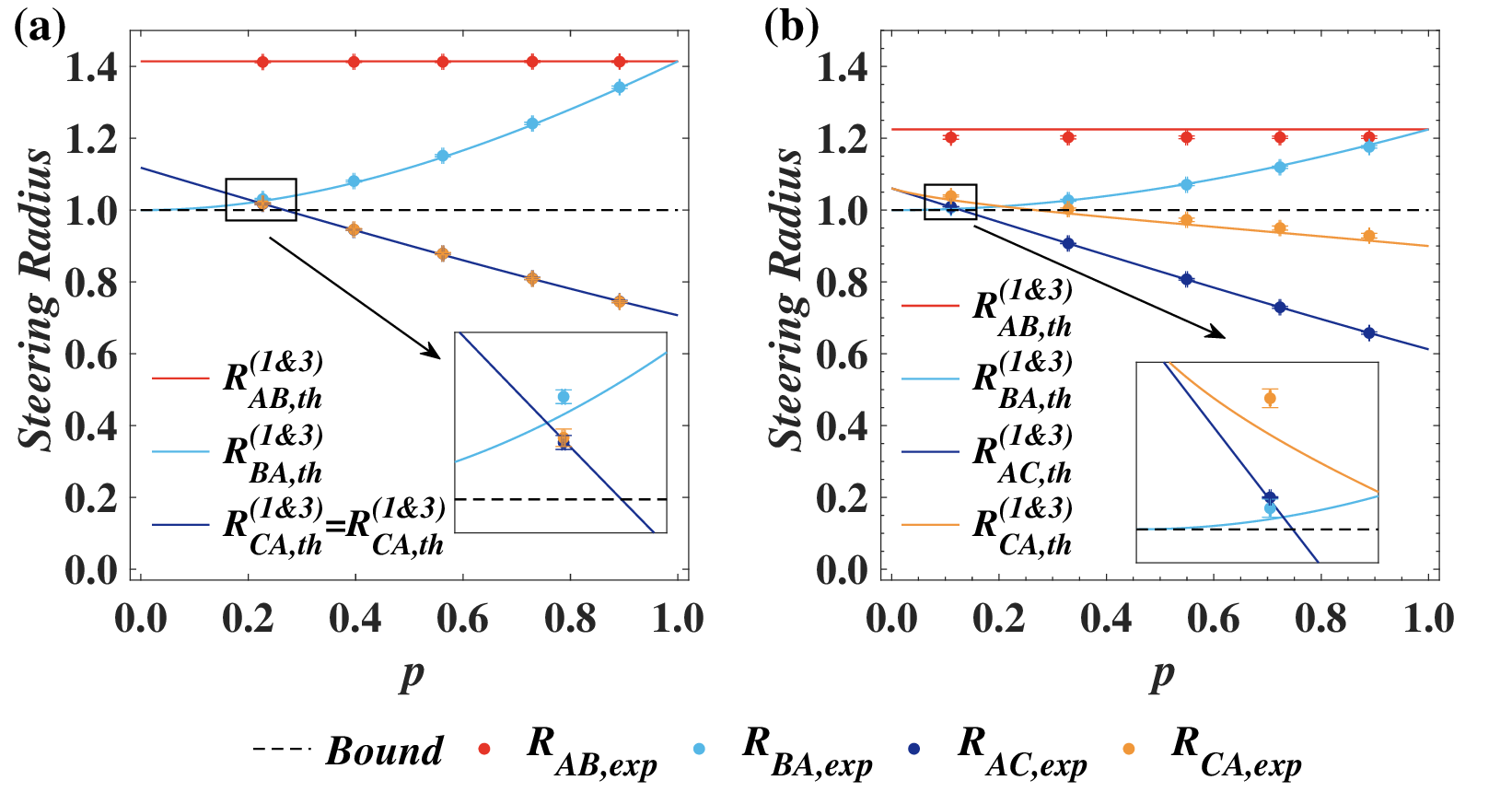}
	\caption{The steering radii as a function of probability $p$ when the measurement strategy is a stochastic combination of case  1 and case 3. (a). The results for a symmetric state with $\{W=0.995\pm0.001, \theta=\pi/4\}$.  (b). The results for an asymmetric state with $\{W=0.992\pm0.001, \theta=\pi/8\}$. Theoretical and experimental results are represented by solid lines and circles, respectively. The dashed black lines denote boundaries between steerable and unsteerable states.  The insets in (a) and (b) provide a magnification of the regions where two-way steerability could be shared among Alice, Bob, and Charlie simultaneously. Error bars are estimated by the Poissonian statistics of two-photon coincidences.}
	\label{result3}
\end{figure}

Finally, we investigate the sequential sharing of steerability by combining two different projective measurement strategies with a probability distribution$\{p(\lambda)\}$. When combining measurement strategies shown in case 1 and case 2, or combining measurement strategies shown in case 2 and case 3, the sharing of two-way steering is absent, as detailed in Appendix I. D and Appendix II. Here, we are mainly interested in the scenario where the measurement strategy is a combination of case 1 and case 3 with $\{p(1), p(3)\} = \{p, 1-p\}$. The steering shareability of a symmetric state with $\{W=0.9955\pm0.0007, \theta=\pi/4\}$, and an asymmetric state with $\{W=0.9931\pm0.0012 , \theta=\pi/8\}$, is shown in FIG. \ref{result3} (a) and (b), respectively. Unlike the previous deterministic measurement strategies, which can only share one-way steering, asymmetric two-way steerability can also be shared among Alice, Bob, and Charlie. For symmetric states, the sharing of asymmetric two-way steering appears when $p\in(0,2 \!-\!\sqrt{3})$, which is observed experimentally when $p=0.2270\pm0.0012$ with $R_{AB}=1.4124\pm0.0024$, $R_{BA}=1.0307\pm0.0021$, $R_{AC}= 1.0170\pm0.0021$, and $R_{CA}=1.0184\pm0.0026$. However, this range decreases to $p\in(0,2-\sqrt{7/2})$ for asymmetric states. Experimentally, we obtain $R_{AB}=1.2023\pm0.0051$,  $R_{BA}=1.0063\pm0.0027$, $R_{AC}=1.0096\pm0.0002$, and $R_{CA}=1.0392\pm0.0028$ when $p=0.1105\pm0.0007$. All of these steering radii exceed the classical bound and are not equal to each other.

\section{Scalability}
It is worth noting that with our projective measurement strategy, the number of parties that can share two-way steering can be increased to four. For example, Alice measures $ A_{0} \!=\! \sigma_{x}$ and  $A_{1} \!=\! \sigma_{z} $, Charlie measures $ C_{0} \!=\! \sigma_{x}$, $C_{1} \!=\! \sigma_{z} $. While Bob$_1$ (Bob$_2$) stochastically combining measurement strategies shown in case 1 and case 3 with the local probability distribution $\{p(\lambda_1)\} \!=\! \{p_1,1 \!-\! p_1\}$ ($\{p(\lambda_2)\} \!=\! \{p_2, 1 \!-\! p_2\}$). Our numerical simulation results indicate that, if the initial state is maximally entangled, two-way steerability can be sequentially shared among Alice, Bob$_1$, Bob$_2$ and Charlie when $p_1\in(0,0.000978)$ and $p_2\in(\sqrt{4 p_1-p_1^2},(2p_1-0.127)/(2-p_1))$. Moreover, if the local randomness of Bob$_i$ is permitted to be shared between Alice and Bob$_i$, the triple sharing of steerability can be improved.  However, Choi \textit{et al.} employ three-measurement settings to obtain the same number of steerability sharing parties as ours \cite{Choi2020}.

Furthermore, our projective measurement strategy enables the unbounded sharing of one-way steering by using deterministic measurement strategies shown in case 2 or case 3. Suppose there are $n$ Bobs located sequentially between Alice and Charlie, as shown in  FIG. \ref{model}. In case 2, each Bob$_i$ employs two identity projections. It is evident that the state shared between Alice and  Bob$_i$  ($i \in \{1,2,\cdots,n\}$) is the same as the initial state $\rho_{in}$. If the initial state is steerable from Alice to Bob$_1$, Alice can simultaneously steer the states of all sequential Bob$_i$. As for case 3, each Bob$_i$ employs an identity projection and a basis  projection. The state $\rho_{AB_{i+1}}$ shared between Alice and Bob$_{i+1}$ can be obtained by Eq. (\ref{luders}). Consider the simplest scenario where the initial state is a pure two-qubit state with $W=1$. It is easy to demonstrate that  the steering radius $R_{AB_{i+1}}$ from Alice to  Bob$_{i+1}$ remains 1, while the steering radius $R_{B_{i+1}A}$ from Bob$_{i+1}$ to Alice equals $\sqrt{1+\sin^2{(2\theta)}/4^{i}}$, which is larger than 1 if $ \theta\neq 0 $. Obviously, any pure two-qubit entangled state enables the sequential sharing of one-way steering from a single Alice to an arbitrary number of Bobs by using deterministic measurement strategies shown in case 3.

The detailed analysis of sharing one-way steerability and two-way steerability among more than three parties can be found in Appendix. III.

\section{Conclusions and Discussion}
\vspace{-1em}
In summary, we investigated the sharing of EPR steering using projective measurement  protocols and the steering radius criterion. Taking three parties and two-measurement settings as an example, our experimental results show that the one-way steering of a two-qubit entangled state can be shared among Alice, Bob and Charlie when Alice and Charlie employ two basis projections and Bob employs two identity projections or one basis projection and the other identity projection. Additionally, compared to when the initial state is symmetric, more direction-related steering classes emerge when the initial state is asymmetric. Furthermore, the sharing of asymmetric two-way steering can also be observed when Bob is permitted to utilize local randomness to combine different measurement strategies. We also demonstrated that our projective measurement strategies enable the unbounded sharing of one-way, even when Bob does not share any local randomness. By leveraging only local randomness, four parties can simultaneously share two-way steering. Our sharing protocols eliminate the need for entanglement assistance in previous unsharp-measurement-based protocols \cite {Sasmal2018,Choi2020} and global randomness assistance in previous projective-measurement-based protocol \cite{Steffinlongo2022},  paving the way for the reuse of quantum correlations.

Our work leaves several relevant open problems: (1) It is known that one-sided unsharp measurements and local randomness-assisted projective measurements enable the sharing of nonlocality among an arbitrary number of sequential parties with a single spatially separated party \cite{Mahato2022,Sasmal2023,Brown2020}. However, with the utilization of two-sided unsharp measurements, Bell nonlocality can only be shared among a limited number of parties. This raises the question: How many pairs of parties can share EPR steering when employing two-sided sequential projective measurements? (2) In the case of two-measurement settings, the number of parties that can share two-way steering is limited to four.  How many shared parties can be increased if one increases the number of measurement settings? (3) Multi-particle and high-dimensional EPR steering can give rise to stronger forms of nonlocal correlations compared to qubit systems, offering significant advantages for quantum information processing~\cite{Designolle2021}. How can the projective measurement strategy to share these correlations be constructed and how can the number of correlation-reuse parties be increased? (4) Sequential unsharp-measurement-based protocols have been demonstrated to yield results that are impossible in the standard (non-sequential) scenario, such as unbounded randomness generation, interesting information-disturbance trade-offs, and advantages in communication complexity. However, what advantages do sequential projective-measurement-based protocols obtain? (5) From the perspective of resource theory \cite{Chitambar2019}, it remains unclear whether correlation sharing via sequential measurements consumes fewer resources for quantum information tasks compared to the standard scenario. Furthermore, the actual advantages in sharing steering and other types of quantum correlations remain ambiguous.

\section{Acknowledgments}
This work was supported by the Natural Science Foundation of Shandong Province of China (Grant No. ZR2021ZD19), the Fundamental Research Funds for the Central Universities (Grants No. 202364008), and the Young Talents Project at Ocean University of China (Grant No. 861901013107).

\section*{Data availability}

The data cannot be made publicly available upon publication because they are not available in a format that is sufficiently accessible or reusable by other researchers. The data that support the findings of this study are available upon reasonable request from the authors. \\

\bibliographystyle{apsrev4-2}
\bibliography{ref.bib}

\onecolumngrid
\appendix
\renewcommand\thesection{\Roman{section}}
\setcounter{equation}{0}

\section{Appendix.I Details for calculating steering radius}
\label{app: cal_srd}

In this work, we study the sharing of asymmetric Einstein-Podolsky-Rosen (EPR) steering with projective measurements. To accurately distinguish no-way steering, one-way steering and two-way steering, a parameter known as the steering radius was employed to quantify steerability for each party. Below is the calculation of steering radii among Alice, Bob, and Charlie for a family of two-qubit states $\rho(W,\theta) $, as depicted in Eq. (\ref{state}) in the main text. According to the symmetrical property of the steering ellipsoid of  $\rho(W,\theta) $, the optimal two-measurement setting directions are $\{\vec{x}, \vec{z} \}$. In our EPR steering sharing protocol, there are three deterministic projective measurement strategies:\\
\textbf{Case 1 ($\lambda=1$):} Alice measures $A_{0}=\sigma_{x}$ and $A_{1}=\sigma_{z}$, Bob measures $B^{(1)}_0=\sigma_x$ and $B^{(1)}_1=\sigma_z$, and Charlie measures $C_{0}=\sigma_{x}$ and $C_{1}=\sigma_{z}$;\\
\textbf{Case 2 ($\lambda=2$):} Alice measures $A_{0}=\sigma_{x}$ and $A_{1}=\sigma_{z}$, Bob measures $B^{(2)}_0=B^{(2)}_1=I$, and Charlie measures $C_{0}=\sigma_{x}$ and $C_{1}=\sigma_{z}$;\\ 
\textbf{Case 3 ($\lambda=3$):} Alice measures $A_{0}=\sigma_{x}$ and $A_{1}=\sigma_{z}$, Bob measures $B^{(3)}_0=I$ and $B^{(3)}_1=\sigma_z$, and Charlie measures $C_{0}=\sigma_{x}$ and $C_{1}=\sigma_{z}$.\\   

\subsection{A. Case 1}
After Alice completes her measurements $\{ A_ {a \vert x}\}$, the corresponding probabilities $\{p^B_ {a \vert x}\}$ and Bloch vectors $\{\vec{v}^B_ {a \vert x}\}$ of Bob's normalized conditional states can be expressed as
\begin{equation}
	\begin{cases}
		
		p^B_{0\vert0} = \frac{1}{2},\ \vec{v}^B_{0\vert0} = (W \sin{(2\theta)}, 0, \cos{(2\theta)}),\\
		
		p^B_{1\vert0} = \frac{1}{2},\ \vec{v}^B_{1\vert0} = (-W \sin{(2\theta)}, 0, \cos{(2\theta)}),\\
		
		p^B_{0\vert1} = \frac{1}{2} (1+W\cos{(2\theta)}),\ \vec{v}^B_{0\vert1} = (0, 0, \frac{W + \cos{(2\theta)}}{1 + W \cos{(2\theta)}}),\\
		
		p^B_{0\vert1} = \frac{1}{2} (1-W\cos{(2\theta)}),\ \vec{v}^B_{1\vert1} = (0, 0, \frac{W - \cos{(2\theta)}}{-1 + W \cos{(2\theta)}}).\\
	\end{cases}
\end{equation}
It has been proven that four local hidden states are sufficient to construct these conditional states if a local hidden state model exists \cite{Sun2016}. The constructed relationship can be expressed as
\begin{equation}
	\begin{cases}
		p^B_{0 \vert 0} = p^B_1 + p^B_2,\ p^B_{0 \vert 0}  \vec{v}^B_{0 \vert 0} = p^B_1 \vec{v}^B_1 + p^B_2 \vec{v}^B_2,\\
		p^B_{1 \vert 0} = p^B_3 + p^B_4,\ p^B_{1 \vert 0}  \vec{v}^B_{1 \vert 0}= p^B_3 \vec{v}^B_3 + p^B_4 \vec{v}^B_4,\\
		p^B_{0 \vert 1} = p^B_1 + p^B_4,\ p^B_{0 \vert 1}  \vec{v}^B_{0 \vert 1} = p^B_1 \vec{v}^B_1 + p^B_4 \vec{v}^B_4,\\
		p^B_{1 \vert 1} = p^B_2 + p^B_3,\ p^B_{1 \vert 1}  \vec{v}^B_{1 \vert 1} = p^B_2 \vec{v}^B_2 + p^B_3 \vec{v}^B_3.\\
	\end{cases}
	\label{rcrl}
\end{equation}
${p^B_{i}}$ and ${\vec{v}^B_{i}}$ ($i\in\{1,2,3,4\}$) represent the corresponding probability and Bloch vector of the local hidden state, respectively. The steering radius $ R^{(1)}_{BA} $ from Alice to Bob can be redefined as  $
R^{(1)}_{BA} =  \min \limits_{\lbrace   p^B_{i}\rho^B_{i}\rbrace} \lbrace {\rm max} \lbrace \vert  \vec{v}^B_{i} \vert \rbrace \rbrace \rbrace  $. According to this definition, the optimal solution of Eq. (\ref{rcrl}) can be obtained when the lengths of the Bloch vectors of the four hidden states are equal. Thus, $R^{(1)}_{AB}  $ can be expressed as
\begin{equation}
	{R^{(1)}_{AB}} = \vert\vec{v}^B_1\vert =\frac{\sqrt{\tan ^2(2 \theta ) \left[4 (W \cos (2 \theta )+1)^2-\sqrt{T_1}\right]^2+64 (\cos (2 \theta )+W)^2}}{8 (W \cos (2 \theta )+1)},
	\label{Eq.A7}
\end{equation}
where $T_1 =2 W^4 \cos (8 \theta )+8 \left(W^4-6 W^2+4\right) \cos (4 \theta )+6 \left(W^4-8 W^2+8\right)$. After Bob completes his measurements $\{ B_ {b \vert y}\}$, the corresponding probabilities $\{p^A_ {b \vert y}\}$ and Bloch vectors $\{\vec{v}^A_ {b \vert y}\}$ of Alice's normalized conditional states can be expressed as 
\begin{equation}
	\begin{cases}
		p^A_{0\vert0} = \frac{1}{2},\ \vec{v}^A_{0\vert0} = (W \sin{(2\theta)}, 0, W \cos{(2\theta)}),\\
		
		p^A_{1\vert0} = \frac{1}{2}, \ \vec{v}^A_{1\vert0} = (-W \sin{(2\theta)}, 0, W \cos{(2\theta)}),\\
		
		p^A_{0\vert1} = \cos^2{\theta},\ \vec{v}^A_{0\vert1} = (0, 0, W),\\
		
		p^A_{1\vert1} = \sin^2{\theta},\ \vec{v}^A_{1\vert1} = (0, 0, -W).\\
	\end{cases}
\end{equation}
Using a method similar to that shown above, we can obtain the steering radius $ R^{(1)}_{BA} $ from Bob to Alice, which can be expressed as
\begin{equation}
	R^{(1)}_{BA} = \vert\vec{v}^A_1\vert =W\sqrt{1+\sin^2{(2\theta)}}.
	\label{Eq.A6}
\end{equation}

When Bob finishes all his measurements, the state  $\rho^{(1)}_{AC}$ shared between Alice and Charlie can be expressed as 

\begin{equation}
	\rho^{(1)}_{AC }= \frac{1}{8}
	\left(
	\begin{array}{cccc}
		\Omega _+ \!+\! \omega _+ \! \cos (2 \theta ) & 0 & 0 & W \! \sin (2 \theta ) \\
		0 & \Omega _- \!-\! \omega _- \! \cos (2 \theta ) & W \! \sin (2 \theta ) & 0 \\
		0 & W \! \sin (2 \theta ) & \Omega _- \!+\! \omega _- \! \cos (2 \theta ) & 0 \\
		W \! \sin (2 \theta ) & 0 & 0 & \Omega _+ \!-\! \omega _+ \! \cos (2 \theta ) \\
	\end{array}
	\right),
\end{equation}
where $\Omega_{\pm} \!=\! 2 \!\pm\! W$ and $\omega_{\pm} \!=\! 1 \!\pm\! 2W$. Similarly, after Alice (Charlie) performs all measurements on $\rho^{(1)}_{AC}$, we can obtain the corresponding probabilities and Bloch vectors of Charlie's (Alice's) normalized conditional states. Then the steering radius $ R^{(1)}_{AC} $ ($ R^{(1)}_{CA} $) from Alice (Charlie) to Charlie (Alice) can be also obtained: 
\begin{equation}
	{R^{(1)}_{AC}} = \frac{\sqrt{\tan ^2(2 \theta ) \left[4 (W \cos (2 \theta )+1)^2-\sqrt{T_1}\right]^2+64 (\cos (2 \theta )+W)^2}}{16 (W \cos (2 \theta )+1)},
	\label{Eq.A9}
\end{equation}

\begin{equation}
	{R^{(1)}_{CA}} = \frac{W \sqrt{256 (2 \cos (2 \theta )+1)^2+J^2 \tan ^2(2 \theta )}}{16 (\cos (2 \theta )+2)},
	\label{Eq.A8}
\end{equation}
where $J = 16 \cos (2 \theta )+2 \cos (4 \theta )-\sqrt{2 (164 \cos (4 \theta )+\cos (8 \theta )+291)}+18 $. In this case, $R^{(1)}_{AB} \in [0,\sqrt{2}]$, $R^{(1)}_{BA} \in [0, \sqrt{2}]$, $R^{(1)}_{AC} \in [0,1/\sqrt{2}]$ and $R^{(1)}_{CA} \in [0, 1]$. Clearly, the steering between Alice and Bob is possible whereas that between Alice and Charlie is impossible. 

\subsection{B. Case 2}
Since Bob employs two identity projections, the state shared between Alice and Charlie is same as that shared between Alice and Bob, i.e., $\rho^{(2)}_{AC}=\rho(W,\theta) $. Additionally, the measurements performed by Alice and Charlie in this case are the same as those performed by Alice and Bob in case 1. Obviously, $ R^{(2)}_{AB}  =R^{(2)}_{AC}=R^{(1)}_{AB} $ and $ R^{(2)}_{CA}=R^{(1)}_{BA} $. $R^{(2)}_{AB} $  is the steering radius from Alice to Bob in case 2. $ R^{(2)}_{AC} $ and $ R^{(2)}_{CA} $ are defined in a similar way. The steerability of Alice and Charlie is equal to that of Alice and Bob in case 1, respectively. When Bob finishes all his measurements, it is easy to demonstrate that Alice's four conditional states are the same, which can be expressed as $\rho_A=\mathrm{Tr}_B[\rho_{AB}]$. The corresponding Bloch vector is  $\vec{v}^A = (0, 0, W\cos{(2\theta)})$. And the steering radius from Bob to Alice can be expressed as $R^{(2)}_{BA} = \vert\vec{v}^A\vert = W \cos{(2\theta)}$. It is obvious that $R^{(2)}_{AB} = R^{(2)}_{AC} \in [0, \sqrt{2}]$, $R^{(2)}_{BA} \in [0, 1]$ and $R^{(2)}_{CA} \in [0, \sqrt{2}]$. Thus, only steering from Bob to Alice is forbidden in case 2.

\subsection{C. Case 3}
Alice performs the same projective measurements as those in case 1 and the conditional states of Bob  are also identical to those of Bob in case 1, it is evident that the steering radius $ R^{(3)}_{AB}$ from Alice to Bob equals to $  R^{(1)}_{AB}$. When Bob measures $B^{(3)}_0 = I$, Alice only obtains one conditional state which is same as that in case 2. When Bob measures $B^{(3)}_1$, Alice obtains two conditional states identical to those in case 1. After Bob completes all his measurements, the corresponding probabilities and Bloch vectors of Alice's normalized conditional states can be expressed as 
\begin{equation}
	\begin{cases}
		p^A_{0\vert0} = 1,\ \vec{v}^A_{0\vert0} = (0, 0, W \cos{(2\theta)}),\\
		p^A_{0\vert1} = \cos^2{\theta},\ \vec{v}^A_{0\vert1} = (0, 0, W),\\
		p^A_{1\vert1} = \sin^2{\theta},\ \vec{v}^A_{1\vert1} = (0, 0, -W). 
	\end{cases}
\end{equation}
Clearly, these conditional states are collinear and located along the z-axis with $\vec{v}^A_{0\vert0}$ lying between $\vec{v}^A_{0\vert1}$ and $\vec{v}^A_{1\vert1}$. Two states  $\{\rho^A_1,\rho^A_2 \}=\{ \rho^A_{0\vert1},\rho^A_{1\vert1}\}$,  with the probability distribution $\{p^A_1,  p^A_2\} = \{\cos^2{\theta}, \sin^2{\theta} \}$, are sufficient to construct these conditional states. The steering radius $ R^{(3)}_{AB}$ from Bob to Alice expressed as $R^{(3)}_{BA} = \vert\vec{v}^A_{0\vert1}\vert=W $. Obviously,  $R^{(3)}_{BA} \in [0,1]$, steering from Bob to Alice is forbidden.

\quad

When Bob finishes all his measurements, the state $\rho^{(3)}_{AC}$ shared between Alice and Charlie can be expressed as 

\begin{equation}
	\rho^{(3)}_{AC }=\frac{1}{2} \left(
	\begin{array}{cccc}
		(1+W) \cos ^2(\theta ) & 0 & 0 & W \sin (\theta)\cos(\theta) \\
		0 & (1-W) \sin ^2(\theta ) & 0 & 0 \\
		0 & 0 & (1-W) \cos ^2(\theta ) & 0 \\
		W \sin ( \theta ) \cos ( \theta ) & 0 & 0 & (1+W) \sin ^2(\theta ) \\
	\end{array}
	\right).
	\label{Eq.A7 }
\end{equation}
Similarly, after Alice (Charlie) performs all measurements on $\rho^{(3)}_{AC}$, we can obtain the corresponding probabilities and Bloch vectors of Charlie's (Alice's) normalized conditional states. Then the steering radius $ R^{(3)}_{AC} $ ($ R^{(3)}_{CA} $) from Alice (Charlie) to Charlie (Alice) can be also obtained, which can be expressed as
\begin{equation}
	{R^{(3)}_{AC}} =\frac{\sqrt{\tan ^2(2 \theta ) \left[4 (W \cos (2 \theta )+1)^2-\sqrt{K}\right]^2+256 (\cos (2 \theta )+W)^2}}{16 (W \cos (2 \theta )+1)},
\end{equation}
\begin{equation}
	R^{(3)}_{CA} = W \sqrt{1+\frac{\sin^2{(2\theta)}}{4}},
\end{equation}
where $K \!=\! 2W^4 \! \cos (8\theta)+8 \left(W^4 \!-\! 18W^2 \!+\! 16\right) \! \cos(4\theta) \!+\! 6\left(W^4 \!-\! 24W^2 \!+\! 24\right)$. In this case, it can be deduced that $R^{(3)}_{AB} \in [0, \sqrt{2}]$, $R^{(3)}_{BA} \in [0, 1]$, $R^{(3)}_{AC} \in [0, \sqrt{5}/2]$ and $R^{(3)}_{CA} \in [0, \sqrt{5}/2]$, making the steering from Bob to Alice unattainable.

\subsection{D. Stochastically combining the above three cases}
\label{app: calc_srd_combi}
Without loss of generality, let's suppose the classical probability distribution of the above three deterministic measurement strategies is $\{ p(1), p(2), p(3)\} = \{p, q, 1-p-q\}$ with $ 0 \le p \le 1$ and $0 \le q  \le 1-p$.   Since Alice's local measurements and the initial states shared between Alice and Bob are the same in each deterministic measurement strategy, it's evident that the steering radius from Alice to Bob in the stochastically combined measurement strategy can be expressed as $R_{AB}=R^{(1)}_{AB}=R^{(2)}_{AB}=R^{(3)}_{AB}$. The steering radius $R_{BA}$ from Bob to Alice can be obtained by solving the corresponding reconstructed relationship, which can be expressed as
\begin{equation}
	R_{BA} = W \sqrt{\sin^2(\theta ) \sec^2(2 \theta ) (2 p \cos^3(\theta )-\sqrt{H} \sin (\theta ))^2+\left(1-2 q \sin^2(\theta )\right)^2},
\end{equation}
where $H=p^2 \sin ^2(2 \theta )+4q(1-q)\cos ^2(2 \theta )$.

The state shared between Alice and Charlie can be expressed as $\rho_{AC} = \sum_{\lambda} p(\lambda) \rho^{(\lambda)}_{AC}$. Similarly, after Alice (Charlie) performs all measurements on $\rho_{AC}$, we can obtain the corresponding probabilities and Bloch vectors of Charlie's (Alice's) normalized conditional states. Then the steering radius $ R_{AC} $ ($ R_{CA} $) from Alice (Charlie) to Charlie (Alice) can be also obtained, which can be expressed as
\begin{equation}
	R_{AC}=\frac{\sqrt{4 (2-p)^2 (\cos (2 \theta )+W)^2+\tan ^2(2 \theta ) \left[(1+q) (W \cos (2 \theta )+1)^2-\sqrt{Q}\right]^2}}{4 (W \cos (2 \theta )+1)},
\end{equation}

\begin{equation}
	R_{CA} =\frac{W }{4 L}\sqrt{\tan ^2(2 \theta ) \left[L (1+q)-\sqrt{S}\right]^2+64 (2-p)^2 \left(p-4 \cos ^2(\theta )\right)^2},
\end{equation}
where
\begin{equation}
	\begin{split}
		Q =  (1+q)^2 \left(1-W^2 \cos ^2(2 \theta )\right)^2+4 (2-p)^2 \left(1-W^2\right) \cos ^2(2 \theta ),
	\end{split}
\end{equation}

\begin{equation}
	\begin{split}
		L = 2 + (2-p) \cos (2 \theta ),
	\end{split}
\end{equation}

\begin{equation}
	\begin{split}
		{\rm and} \quad  S = 
		&16 (1+q)^2 + (2-p)^4 (1+q)^2-8 (2-p)^2 \left[2 p (p-4) +(1+q)^2\right]\\
		&+8 p (4-p) (2-p)^2 \left(q^2+2 q-7\right) \cos ^2(\theta )-32 (2-p)^4 (1+q)^2 \cos ^6(\theta )\\
		&+ \left[24 (2-p)^4 (1+q)^2-32 (2-p)^2 \left(2 (p-4) p+(1+q)^2\right)\right] \cos ^4(\theta )\\
		&+16 (2-p)^4 (1+q)^2 \cos ^8(\theta ).
	\end{split}
\end{equation}
When combining measurement strategies shown in case 2 and case 3 with $\{p(2), p(3) \} = \{p, 1 \!-\! p \}$, the steering radius from Bob to Alice can be simplified to $R^{(2\&3)}_{BA} = W \sqrt{1 \!-\! p \sin^2(2\theta)} $. Obviously, $R^{(2\&3)}_{BA} \!\in\! [0,1]$,  making the steering from Bob to Alice impossible, and thus, the sequential shearing of two-way steering cannot be achieved.

\section{Appendix II. More experimental results} \label{app: more_result}
\begin{figure}[ht]
	\centering
	\includegraphics[width=\linewidth]{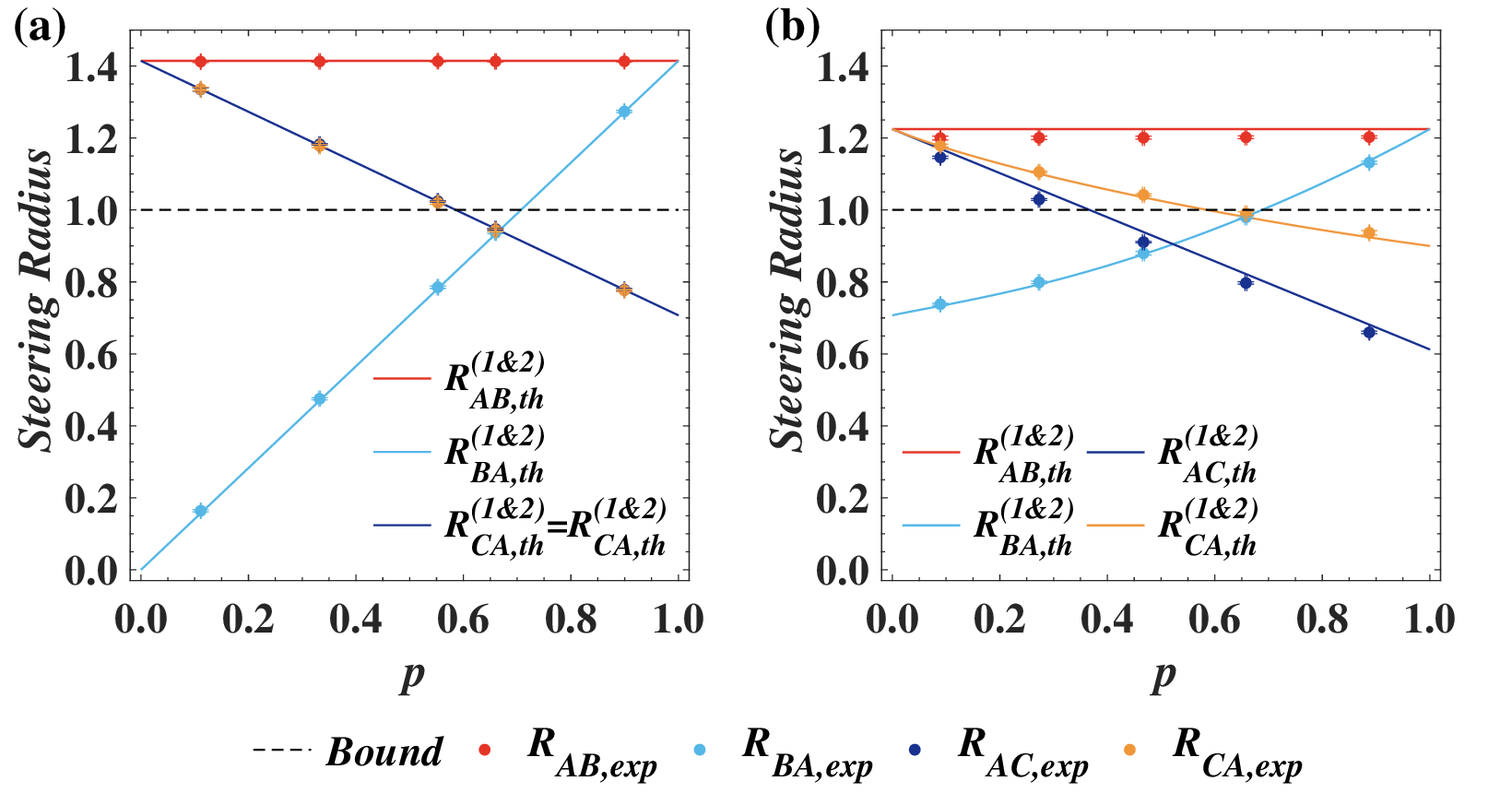}
	\caption{The steering radii as a function of probability $p$ when the measurement strategy is a stochastic combination of case  1 and case 2. (a). The results for a symmetric state with $\{W=0.9955\pm0.0007, \theta=\pi/4\}$.  (b). The results for an asymmetric state with $\{W=0.9931\pm0.0012, \theta=\pi/8\}$ . Theoretical and experimental results are represented by solid lines and circles, respectively. The dashed black lines denote boundaries between steerable and unsteerable states. Error bars are estimated by the Poissonian statistics of two-photon coincidences.}
	\label{result4}
\end{figure}

FIG. \ref{result4} (a) and (b) present the steering radii $\{R^{(1\&2)}_{AB},R^{(1\&2)}_{BA},R^{(1\&2)}_{AC},R^{(1\&2)}_{CA} \}$ as a function of probability $ p$ when the measurement strategy is a combination of case 1 and case 2 with $\{p(1), p(2)\}=\{p,1-p\}$. The steering shareability of a symmetric state with $\{W=0.9955\pm0.0007, \theta=\pi/4\}$ and an asymmetric state with $\{W=0.9931\pm0.0012, \theta=\pi/8\}$  is shown in FIG. \ref{result4} (a) and (b), respectively. Clearly, the sharing of one-way steering is observed whereas that of two-way steering is absent. 

\section{Appendix III. Sharing steerability among more than three parties}  \label{app: unbound}

\subsection{A. Triple sharing of two-way steering with randomness-assisted projective measurement strategy}

We provide a simple example demonstrating that, using local randomness-assisted projective measurements, the sharing of two-way steering among three sequential parties and one spatially separated party is possible. Let the initial state be a pure maximally entangled state, represented by  $\ket{\phi^{+}} = [\ket{HH} + \ket{VV}]/\sqrt{2}$. One qubit of the entangled state is sent to Alice, while the other is sequentially sent to Bob$_1$, Bob$_2$, and Charlie. Suppose Bob$_1$ (Bob$_2$) adopts a stochastically combined measurement strategy with $\{p(\lambda_1)\}=\{p_1,q_1,1-p_1-q_1\}$ ($\{p(\lambda_2)\}=\{p_2,q_2,1-p_2-q_2\}$). $p_1$ and $q_1$ ($p_2$ and $q_2$) are constrained by $0 \le p_1 \le 1$ and $0 \le q_1 \le 1 - p_1$ ($0 \le p_2 \le 1$ and $0 \le q_2 \le 1 - p_2$).  When Alice measures  $ A_{0}=\sigma_{x}$ and  $ A_{1} = \sigma_{z}$, Charlie measures $C_{0} = \sigma_{x}$ and  $C_{1}=\sigma_{z} $, the steering radii $\{R_{AB_1}, R_{B_1A}, R_{AB_2}, R_{B_2A}, R_{AC}, R_{CA}\}$ can be expressed as  
\begin{equation}
	\label{Eq.A18}
	\begin{split}
		&R_{AB_1} =\sqrt{2},\\
		&R_{B_1A} = \sqrt{p_1^2+(1-q_1)^2},\\
		&R_{AB_2} =\frac{1}{2} \sqrt{(2-p_1)^2 + (1-q_2)^2},\\
		&R_{B_2A} = \frac{1}{2} \sqrt{(2-p_1)^2 (1-q_2)^2+p_2^2 (q_1+1)^2}\\
		&R_{AC} = R_{CA} = \frac{1}{4} \sqrt{(2-p_1)^2(2-p_2)^2 + (1+q_1)^2(1+q_2)^2}.\\
	\end{split} 
\end{equation}
By numerically searching the solutions to the inequalities $\{R_{AB_1}>1, R_{B_1A}>1,R_{AB_2}>1, R_{B_2A}1, R_{AC}>1, R_{CA}>1\}$ with the constraint conditions $\{0 \le p_1 \le 1, 0 \le q_1 \le 1-p_1, 0 \le p_2 \le 1, 0 \le q_2 \le 1-p_2\}$, we find that $q_1=q_2=0$, $p_1 \in (0,\ 0.000978)$ and $ p_2 \in (\sqrt{4 p_1-p_1^2},\ (2p_1-0.127)/(2-p_1))$. As an example, taking $p_1 = 0.000097$ and $p_2=0.045$, the corresponding values of the steering radii are  $R_{AB_1} = 1.4142135$, $R_{B_1A} = 1.0000005$, $R_{AB_2} = 1.1176002$, $R_{B_2A} = 1.0000033$ and  $R_{AC} = R_{CA} = 1.0000332$. Clearly, by stochastically combining the deterministic measurement strategies shown in case 1 and case 3, two-way steerability can be simultaneously shared among Alice, Bob$_1$, Bob$_2$, and Charlie.

It should be noting that, if after each round of measurement, each Bob$_i$ is further allowed to inform Alice about the measurement strategy he adopted, the shareability of two-way steering can indeed be further increased.  In this scenario, once each Bob$_i$ has completed a measurement, Alice can determine which measurement strategy the corresponding conditional state she obtains belongs to. The steering radius  $R_{B_iA}$ from Bob$_i$ to Alice can be expressed as a convex combination of those in the corresponding measurement strategies, i.e., $R_{B_iA} = \sum_{\lambda_i} p(\lambda_i) R^{(\lambda_i)}_{B_iA}$. The above steering radii $\{R_{AB_1}, R_{B_1A}, R_{AB_2}, R_{B_2A}, R_{AC}, R_{CA}\}$ can be rewritten as    
\begin{equation}
	\begin{split}
		&R_{AB_1} =\sqrt{2},\\
		&R_{B_1A} = p_1 \sqrt{2} + (1-p_1-q_1),\\
		&R_{AB_2} = \frac{1}{2} \sqrt{(2-p_1)^2 + (1 + q_1)^2},\\
		&R_{B_2A} = \frac{p_2}{2} \sqrt{(2-p_1)^2 + (1 + q_1)^2} + (1-p_2-q_2)(1-\frac{p_1}{2}),\\
		&R_{AB_3} = R_{B_3A} = \frac{1}{4} \sqrt{(2-p_1)^2(2-p_2)^2 + (1+q_1)^2(1+q_2)^2}.\\
	\end{split}
	\label{Eq.A19}
\end{equation}
By numerically searching the solutions to the inequalities $\{R_{AB_1}\!>\!1,R_{B_1A}\!>\!1,R_{AB_2}\!>\!1,R_{B_2A}\!>\!1,R_{AC}\!>\!1,R_{CA}\!>\!1\}$ with the constraint conditions $\{0 \! \le \! p_1 \! \le \! 1, 0 \! \le \! q_1 \! \le \! 1 \!-\! p_1, 0 \! \le \! p_2 \! \le \! 1, 0 \! \le \! q_2 \! \le \! 1 \!-\! p_2\}$, we find that  the triple sharing of two-way steering can be achieved when $q_1=q_2=0$, $p_1 \in (0, 0.0122)$ and $p_2 \in (p_1(2-p_1+\sqrt{5+p_1(p_1-4)}), (2p_1-0.127)/(2-p_1))$. As an example, when $p_1=0.0009$ and $p_2=0.06$, the values of $ R_{AB_1}$, $R_{B_1A}$, $R_{AB_2}$, $R_{B_2A}$, $R_{AC}$ and $R_{CA}$ respectively change to $R_{AB_1} = 1.4142136$, $R_{B_1A} = 1.0003728$, $R_{AB_2} = 1.1176315$, $R_{B_2A} = 1.0066348$ and  $R_{AC} = R_{CA} = 1.0012759$. Clearly, the shareability of two-way steering can indeed be further improved.

\subsection{B. Unbounded sharing of one-way steering with deterministic projective measurement strategy}
\label{sec: unbounded}
We now go beyond the scenario of three sequential parties and show that, by employing deterministic measurement strategies shown in case 1 and case 3, unbounded sharing of one-way steering is possible. Suppose there are $n$ Bobs located sequentially between Alice and Charlie. 

In case 2, each Bob and Charlie employs two identity projections, the state shared between Alice and Bob$_i$ ($i \in \{1,2,\cdots,n\}$) is the same as the initial state $\rho_{in}$. When Bob$_i$ (Charlie) finishes all his measurements,  Alice's four conditional states are the same. The corresponding Bloch vector is same as that of Alice's reduced state, which can be expressed as $\vec{v}^A = (0, 0, W \cos{(2\theta)})$. And the steering radius from Bob$_i$ to Alice can be expressed as $R^{(2)}_{B_{i}A} = \vert\vec{v}^A\vert = W \cos{(2\theta)}$. Obviously, $R^{(2)}_{B_{i} A} \in [0,1]$, the steering from any Bob$_i$ to Alice is impossible. It is easy to demonstrate that the steering radius $R_{AB_{i}}$ from Alice to any Bob$_{i}$ can be expressed as
\begin{equation}
	{R^{(1)}_{AB_{i}}} = \vert\vec{v}^{B_i}_1\vert =\frac{\sqrt{\tan ^2(2 \theta ) \left[4 (W \cos (2 \theta )+1)^2-\sqrt{T_1}\right]^2+64 (\cos (2 \theta )+W)^2}}{8 (W \cos (2 \theta )+1)},
\end{equation}
where $T_1 = 2 W^4 \cos (8 \theta )+8 \left(W^4-6 W^2+4\right) \cos (4 \theta )+6 \left(W^4-8 W^2+8\right)$. Clearly, $ R^{(1)}_{AB_{i}}$ is larger than 1 when $W \in (1/2,1]$ for any $i \in \{1,2,\cdots,n\}$,  which means Alice can steer the states of all Bob$_{i}$ simultaneously.

If each Bob employs a deterministic measurement strategy shown in case 3, the state shared between Alice and Bob$_i$ can be expressed as

\begin{equation}
	\rho_{AB_i} = \frac{1}{2}
	\begin{pmatrix}
		(1+W)\cos^2{\theta} & 0 & 0 & 2^{1-i} W \sin{(2\theta)} \\
		0 & (1-W)\sin^2{\theta} & 0 & 0 \\
		0 & 0 & (1-W)\cos^2{\theta} & 0 \\
		2^{1-i} W \sin{(2\theta)} & 0 & 0 & (1+W)\sin^2{\theta}
	\end{pmatrix}.
\end{equation}
The steering radius $ R^{(3)}_{AB_i } $ from  Alice to Bob$_i$ can be expressed as 
\begin{equation}
	R^{(3  )}_{AB_i} = \frac{ \sqrt{\tan^2{(2\theta)\left[4(W\cos{(2\theta)}+1)^2-\sqrt{T_i}\right]^2} + 4^{i+2}(W+\cos{\theta})^2}}{2^{i+2}(W\cos{(2\theta)}+1)},	\label{Eq.A21}
\end{equation}
where $T_i = -4^{i+2}(W^2-1)\cos^2{(2\theta)}+4(-2+W^2+W^2\cos{(4\theta)})^2$. While the  steering radius $ R^{(3)}_{AB_i} $ from   Bob$_i$ to Alice can be expressed as 
\begin{equation}
	R^{(3)}_{B_iA} = W. \label{Eq.A22}
\end{equation}
Consider the most straightforward scenario where the initial state is a pure two-qubit entangled state, in other words, $W=1$. Eq. (\ref{Eq.A21}) and Eq. (\ref{Eq.A22}) are respectively simplified to  $R^{(3)}_{AB_i} = \sqrt{1+ \sin^2{(2\theta)}/4^{i-1}}$ and $R^{(3)}_{B_iA} = 1$.
Obviously,	$R_{AB_i}$  is larger than 1 if $ \theta\neq 0 $, and $R_{B_iA}$ remains 1. Thus, with the deterministic measurement strategies shown in case 3, pure two-qubit entangled states enable the sharing of one-way steering from a single Alice to an unlimited number of Bobs.

\end{document}